# The Proposal of Improved Component Selection Framework

Weam Gaoud Alghabban, M. Rizwan Jameel Qureshi

Faculty of Computing and Information Technology, King Abdulaziz University, Jeddah, Saudi Arabia
weam_ghabban@yahoo.com, anriz@hotmail.com

**Abstract:** Component selection is considered one of hard tasks in Component Based Software Engineering (CBSE). It is difficult to find the optimal component selection. CBSE is an approach that is used to develop a software system from pre-existing software components. Appropriate software component selection plays an important role in CBSE. Many approaches were suggested to solve component selection problem. In this paper the component selection is done by improving the integrated component selection framework by including the pliability metric. Pliability is a flexible measure that assesses software quality in terms of its components quality. The validation of this proposed solution is done through collecting a sample of people who answer an electronic questionnaire that composed of 20 questions. The questionnaire is distributed through social sites such as Twitter, Facebook and emails. The result of the validation showed that using the integrated component selection framework with pliability metric is suitable for component selection.
[Alghabban WG, Qureshi MRJ. **The Proposal of Improved Component Selection Framework.** *Life Sci J* 2013;10(4):3538-3544]. (ISSN:1097-8135). http://www.lifesciencesite.com. 470

## 1. Introduction

Component Based Software Engineering (CBSE) is concerned with selecting and designing components. It is centered on the idea of developing a system from pre-existing components. Designing a system by reusing existing components leads to faster time to market. However, finding the appropriate set of components that satisfy a set of requirements is becoming more difficult and has its challenges. That means the components must include enough information to allow designers to take decisions when selecting among components.

Several researches have addressed these challenges by reducing the search space for component selection problem. One of these researches is integrated framework. According to (Calvert et al., 2011) this framework estimates whether a combination of components satisfy the system performance requirements by using system modeling and statistical analysis in the first phase. The second phase is the component selection phase. The pliability metric was produced (Pande et al., 2013), which measures software quality as a function of individual components quality, is added to the framework in the first phase in order to enhance it.

This paper's organization is as follow: Section 2 provides an overview of the literature. Section 3 defines the problem statement. The proposed solution for the problem is described in Section 4. Section 5 discusses the validation of the proposed solution by means of the questionnaire. Section 6 shows the results of questionnaire and statistical analysis. Finally the conclusion is presented in Section 7.

## 2. Related Work

Component-Based Software Engineering (CBSE) is an approach which concerned with selecting, composing and designing components. CBSE aims to develop software faster with better quality by using existing components. Selecting components from available component set which can satisfies a given set of requirements with low cost plays an important role to enhance the reusability and quality. For that, (Vescan, 2009) aimed to a selection approach that takes into consideration some attributes of components and the evaluation of them. He considered three attributes for the system which are cost, reusability and functionality and he defined metrics for these attributes.

The existing components selection methods do not address specification of functional and non-functional requirements. Component selection decision is important because the integration risks can be solved by the right selection of components. genetic algorithms based approach is proposed to solve the problem of component selection (Dixit and Saxena, 2009).

Software modularity is used in CBSE development to enhance the comprehensibility and flexibility of software systems. A methodology is proposed to perform the optimal selection of software components for CBSE development based on a modified way of measuring the cohesion and coupling of software modules by using genetic algorithm (Kwong et al., 2010). Information and complexity theory can be examined to measure the cohesion and coupling of software modules.

Complex component selection problem can depend on dynamic environment and its characterization use more than one criterion. An approach is proposed (Vescan et al., 2011) to envisage both of these aspects. This approach uses principles of evolutionary computation and multi-objective optimization. First, the problem is formulated as a multiple objective optimization problem having four objectives: used components number, new requirements number, provided interfaces number and the initial requirements number that are not in solution.

Component selection process is not considering an easy task in CBSE. The cost of the component is considered, (Kaur and Mann, 2010), when selecting a component which is calculated on the basis of quality of component.

Many of the previous researches focused on technical details of component selection while the internal management provided by the models is ignored. Also, some of researches have few formal techniques to consider compatibility and assembly of component selection. For that, (Tang et al., 2011) proposed an optimization model to solve the problem of component selection including reusability and compatibility at the same time. This model depends on genetic algorithm and that helps the developers to select components when they are working on multiple applications concurrently.

A comparison of different methods is provided, (Fahmi and Choi, 2009), that already used for components selection.

In CBSE, component selection is considered a main issue. After completing the requirement analysis and design phase of X model, developers start searching for optimal set of components that meet client's requirements and minimize the overall cost. A new algorithm is proposed (Tomar and Gill, 2013) to choose the optimal components from TCR and RCR. This algorithm depends on best-fit strategy and first-fit strategy for searching the components through SCSP and CCSP.

Designing a system by using the existing components can reduces the time to market. But how to select the right components that meet the system requirements become harder job because components must have enough information that enable designers to take decisions when choosing between them. Also there is difficulty for designers to select between many components with similar functionalities but with different performance and quality. For that, (Calvert et al., 2011) proposed a framework for components selection by using simulated annealing algorithm.

Since CBSE includes reuse of components into new software, it aims to enhance software quality by improving functionality, security, cost and maintainability. There is a need to consider many different quality attributes in the final system. For that, (Pande et al., 2013) developed a flexible evaluation model to enable optimal component selection based on different quality metrics of component and cost. This model based on integer programming to maximize the pliability of the overall system by designing a metric called the pliability metric, which enhances component selection.

Right selection components from components set plays important role for system successfully. For that, a comparison table is drawn, as shown in Table 1, to present each component selection methods with its limitations.

## 3. Problem Statement

Selecting pre-existing software components that satisfy the client's requirements plays a critical role. A formal definition of the problem that addresses in this paper is as follow: consider the client's requirements and there is a set of components available for selection between them. Each component can satisfies a group of requirements. The question: "how to find a set of components that satisfy the requirements?" is considered one of the main problems related to the component selection and reuse. The next section describes the proposed solution for this problem.

## 4. The Proposed Solution for Component Selection Problem

Designing a system by reusing existing components can lead to a faster time to market. However, selection the optimum set of components from a component library that satisfies the functional and non-functional system requirements has its challenges. There are several approaches have addressed these challenges. Calvert et al. (2011) developed one approach which is integrated component selection framework.

The system performance cannot be expressed by terms of its individual components' performance. That is because performance cannot be measured for each component in isolation, rather than it can be measured for the integrated system. For that, the framework uses system modeling and analysis, which is the first phase (Figure 1), to conclude the dependencies between system performance and components attributes. In the first phase, the regression analysis computes the dependency of system performance with attributes of individual components and probability analysis computes the probability of how can a certain component satisfy certain performance requirements. The outputs of this phase are regression equations and conditional probabilities which are used in the second phase (Calvert et al., 2011).

The second phase begins from the system requirements and specification. In the component filtering step, the search space is reducing by eliminating the components that do not satisfy the system specification constraints. Then, the selection algorithm is chosen from three algorithms which are greedy, intelligent greedy and simulated annealing algorithms. Each one of these algorithms produce a combination of components. Then, performance of this combination of components is estimated by using the output of phase 1 which is regression equations. Then, this proposed components performance is compared with giving system performance requirements. If it is satisfied the giving system performance requirements, then it will be as a recommended solution for the user. If not, the process loops back again and the selected algorithm generates another combination of components and the process continues (Calvert et al., 2011).

Table 1. Comparison of brief related work

| Paper Title | Limitations |
| --- | --- |
| A Metrics-Based Evolutionary Approach for the Component Selection Problem (Vescan, 2009). | • Limitation is metrics that are used to select component based on three attributes: cost, reusability and functionality.<br>• It is not specifying compatibility between two connected components. |
| Software Component Retrieval Using Genetic Algorithms (Dixit and Saxena, 2009). | • Limitation of this approach is retrieving component is expensive. |
| Optimization of Software Components Selection for Component-Based Software System Development (Kwong et al., 2010). | • Limitation of this methodology is that it includes judgments from software development teams to determine the interaction scores and function ratings. |

| Paper Title | Limitations |
| --- | --- |
| A Hybrid Evolutionary Multiobjective Approach for the Dynamic Component Selection Problem (Vescan et al., 2011). | • This approach limitation is that it deals with simple problems without considering hierarchies of components. |
| Component Selection for Component based Software Engineering (Kaur and Mann, 2010). | • Limitation in this approach is that requirement specification may not detailed enough for evaluating OTSO software alternatives.<br>• Case studies that carried out in this approach were relatively small and the evaluation processes and resulting criteria were extensive. |
| An Optimization model for Software Component Selection under Multiple Applications Development (Tang et al., 2011). | • Limitation of this model does not focus on functional and non functional requirements that should be considered simultaneously.<br>• Customized genetic algorithm could not guarantee the optimal solution. |
| A Study on Software Component Selection Methods (Fahmi and Choi, 2009). | • This approach limitation is that there must be case base and database of components that do not have all cases. |
| Algorithm for Component Selection to Develop Component-Based Software with X Model (Tomar and Gill, 2013). | • Limitation of this approach is time and cost of development are high.<br>• How to use it to develop component based software by choosing optimal set of components from X model repositories that meets client's requirements. |
| An Integrated Component Selection Framework for System-Level Design (Calvert et al., 2011). | • Limitation in this approach is that the number of system requirements used in component selection is small. |
| Optimal Component Selection for Component Based Software Development using Pliability Metric (Pande et al., 2013). | • Limitation of this model is not include more quality metrics for components that are easy to calculate and more feasible to use.<br>• One limitation is that how to devise a formal methodology for determining the relative weights to be assigned to the different quality metrics based on stakeholder input. |

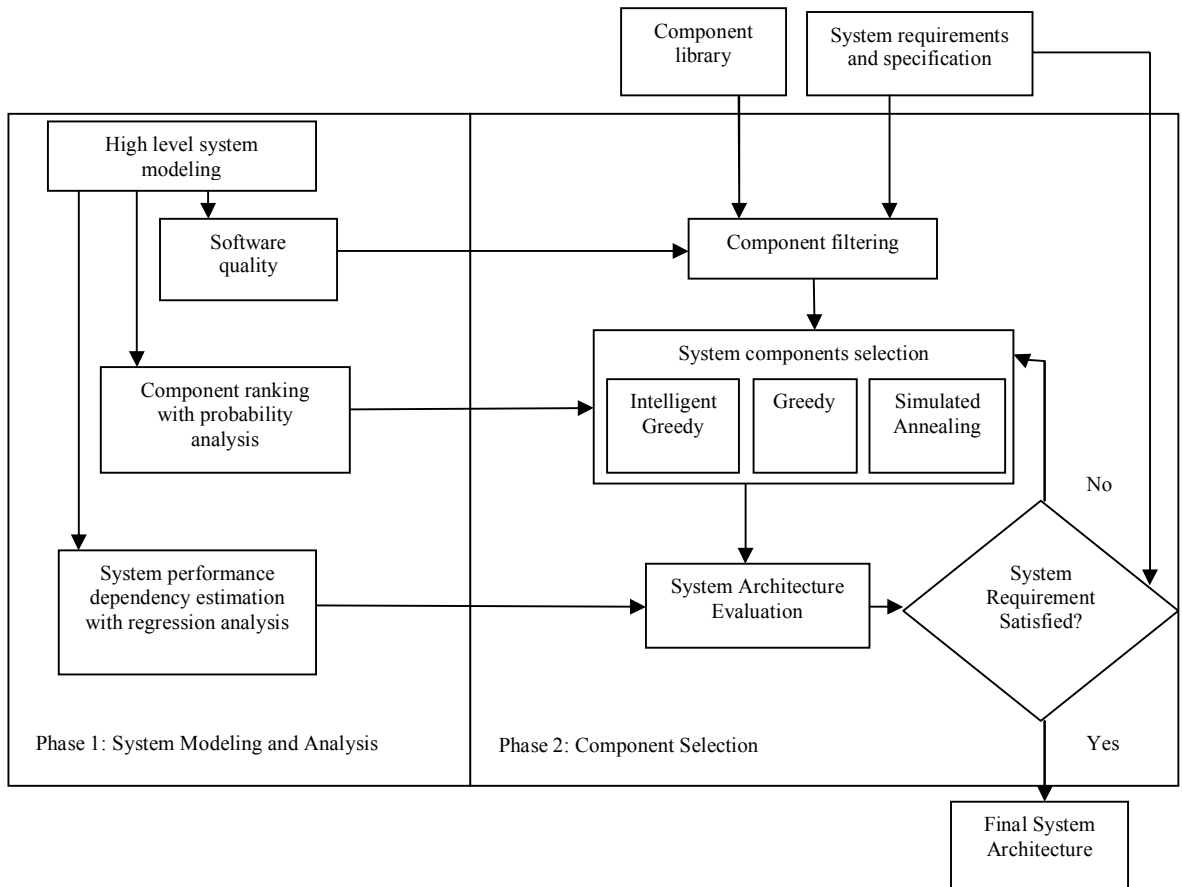

Figure 1. The proposed integrated component selection framework

System performance is considered as an important perspective in the integrated component selection framework. Another robust perspective must be taken into consideration when performing component selection is that the software quality enhancement which is one aim of CBSE. The software quality can be achieved by improving maintainability, security, functionality, cost and other aspects. One important consideration is that many software quality attributes must be considered in the final system. Moreover, these quality attributes have different relative importance which depends on the software system type being developed. So, there is a need for an approach to enable the component selection based on the software quality importance. For that, a flexible metric which is pliability metric is used to evaluate software quality. Pande et al. (2013) produced the pliability metric that measures the software quality as a function of individual components quality while enabling the customer to define and prioritize the quality attributes that are important.

This paper proposes an approach that can solve the components selection problem by integrating the pliability metric to the existing integrated component selection framework to improve its performance and utilize multiple dimensions of quality that enable flexibility for the system being designed. The software quality step is adding in the first phase (Figure 1). The total software quality measure, Q, can be defined based on a set of quality attributes which includes reliability, performance, fault tolerance, security, safety, availability, maintainability and testability. Other quality attributes which consider important can be included (Pande et al., 2013). So, the total software quality can be expressed as weighted linear combination of these attributes values as:

$Q = w_R R + w_P P + w_F F + w_{Sa} Sa + w_{Se} Se + w_{Av} Av + w_T T + w_M M$

where, R= Reliability, P= Performance, F= Fault tolerance, Sa= Safety, Se= Security, Av= Availability, T= Testability, M= Maintainability.

For each one of attributes, a weighted value is assigned and the sum of all weights is equal to 1 as:

$w_R + w_P + w_F + w_{Sa} + w_{Se} + w_{Av} + w_T + w_M = 1$

This metric provides a flexible way to assign weighted values for each attribute depending on software type. For example, in financial system, a high weighted value is assigned for security attribute. In ecommerce system, the performance, reliability, security, maintainability and availability attributes are considered important relative to other attributes. So, high weighted values can be assigned to them.

It is important to normalize quality attribute measures so it allows comparing between them in a right way. For that, a normalized level, (qhi), for each quality attribute (h) contributed by a component (i) is defined. A normalized value from 0 to 10 can be calculated by this formula:

qhi= ((Raw QA (h, i)) / (Max (h)) × 10

which is a ratio raw measured value of component (i) with respect to quality attribute (h) to maximum raw measured value of h attained by any component (i). By using this method, quality attributes measurements are normalized in a meaningful way.

After calculating the total weighted normalized value, this value is used in component filtering step (Figure 1).

## 5. Validation of the Proposed Solution

In this paper the validation of the proposed solution is done by using an electronic questionnaire. The electronic questionnaire composes of 20 close ended questions divided into 3 goals and it is targeted to software engineering specialists. Questions were arranged according to their relevancy to defined goals. This questionnaire is distributed via social sites such as Twitter, Facebook and email. The total number of people who answered is 46 respondents which formed the study sample. The likert scale is used to answer the questionnaire (Table 2).

Table 2. Likert scale

| 1 | Very low |
|---|---|
| 2 | Low |
| 3 | Nominal |
| 4 | High |
| 5 | Very high |

Once data is collected, a statistical analysis is applied on it. The analytic form is represented by using frequency tables and bar charts. The next section shows the results of questionnaire and statistical analysis.

## 6. Findings

This section shows the results of statistical analysis for each goal.

### 6.1 Cumulative Statistical Analysis of Goal 1

Goal 1: need for effective framework/software that automatically selecting component especially when there are a large number of components. This goal covers questions that are related to selecting the right components among a large number of components is considered a major issue in component selection especially when this is done automatically rather than manually. As it is clear from the cumulative descriptive analysis of goal 1 that 1.71% of respondents are very low agreed, 4.85% are low agreed, 22.57 % are neither agree nor disagree, 41.57% of the them are high agreed and 29.28 % of them are very high agreed.

The result of the analysis of the goal 1 is shown in (Table 3).

Table 3. Cumulative statistical analysis of goal 1

| Q. No. | Very low | Low | Nominal | High | Very high |
|---|---|---|---|---|---|
| 1 | 2% | 11% | 24% | 35% | 28% |
| 2 | 0% | 2% | 20% | 48% | 30% |
| 3 | 0% | 2% | 35% | 48% | 15% |
| 4 | 2% | 0% | 7% | 39% | 52% |
| 5 | 2% | 4% | 16% | 35% | 43% |
| 6 | 4% | 11% | 43% | 42% | 0% |
| 7 | 2% | 4% | 13% | 44% | 37% |
| Total | 12% | 34% | 158% | 291% | 205% |
| Average | 1.71% | 4.85% | 22.57% | 41.57% | 29.28% |

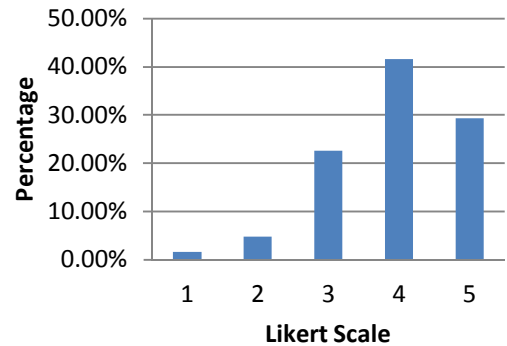

Figure 2. Graphical representation of goal 1

### 6.2 Cumulative Statistical Analysis of Goal 2

Goal 2: the proposed integrated component selection framework can improve and make a perfect selection of components. This goal covers questions that are related to the proposed integrated component selection framework if it can improve a perfect selection of components or not. As it is clear from the cumulative descriptive analysis of goal 2 that 1.87% of respondents are very low agreed, 4.25% are low agreed, 25.25% are neither agreed nor disagreed, 44.87% are high agreed and 23.75% are very high agreed.

The result of the analysis of the goal 2 is shown in (Table 4).

Table 4. Cumulative analysis of goal 2

| Q. No. | Very low | Low | Nominal | High | Very high |
|---|---|---|---|---|---|
| 8 | 2% | 2% | 15% | 59% | 22% |
| 9 | 9% | 0% | 13% | 33% | 45% |
| 10 | 2% | 0% | 9% | 61% | 28% |
| 11 | 0% | 4% | 22% | 37% | 37% |
| 12 | 0% | 4% | 30% | 39% | 27% |
| 13 | 0% | 2% | 39% | 46% | 13% |
| 14 | 2% | 20% | 46% | 30% | 2% |
| 15 | 0% | 2% | 28% | 54% | 16% |
| Total | 15% | 34% | 202% | 359% | 190% |
| Average | 1.87% | 4.25% | 25.25% | 44.87% | 23.75% |

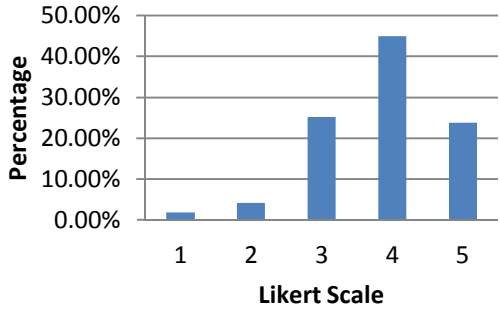

Figure 3. Graphical representation of goal 2

### 6.3 Cumulative Statistical Analysis of Goal 3

Goal 3: user satisfaction by using proposed integrated component selection framework. This goal covers questions that are related to if the proposed integrated component selection framework satisfies the users or not. As it is clear from the cumulative descriptive analysis of goal 3 that 1.6% of respondents are very low agreed, 6.8% are low agreed, 32.6% are neither agreed nor disagreed, 43.6% are high agreed and 15.4% are very high agreed.

The result of the analysis of the goal 3 is shown in (Table 5).

### 6.4 Cumulative Statistical Analysis of 3 Goals

The evaluation of 3 goals together showing that 1.75% are in favor of very low, 5.1% are supporting low, 26.15% are neither agreed nor disagreed, 43.4% are highly agreed and 23.60% are very highly agreed.

Table 5. Cumulative analysis of goal 3

| Q. No. | Very low | Low | Nominal | High | Very high |
|---|---|---|---|---|---|
| 16 | 0% | 4% | 28% | 52% | 16% |
| 17 | 4% | 28% | 46% | 20% | 2% |
| 18 | 0% | 0% | 41% | 48% | 11% |
| 19 | 2% | 2% | 41% | 37% | 18% |
| 20 | 2% | 0% | 7% | 61% | 30% |
| Total | 8% | 34% | 163% | 218% | 77% |
| Average | 1.6% | 6.8% | 32.6% | 43.6% | 15.4% |

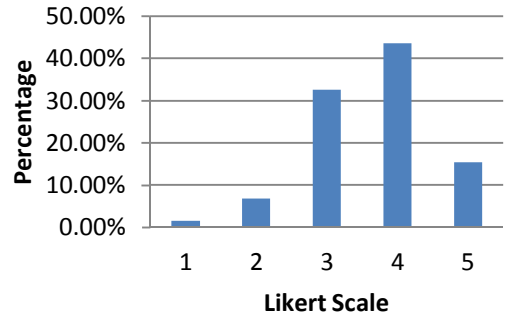

Figure 4. Graphical representation of goal 3

The result of the analysis of the 3 goal is shown in (Table 6).

Table 6. Cumulative analysis of 3 goals

| Q. No. | Very low | Low | Nominal | High | Very high |
|---|---|---|---|---|---|
| 1 | 2% | 11% | 24% | 35% | 28% |
| 2 | 0% | 2% | 20% | 48% | 30% |
| 3 | 0% | 2% | 35% | 48% | 15% |
| 4 | 2% | 0% | 7% | 39% | 52% |
| 5 | 2% | 4% | 16% | 35% | 43% |
| 6 | 4% | 11% | 43% | 42% | 0% |
| 7 | 2% | 4% | 13% | 44% | 37% |
| 8 | 2% | 2% | 15% | 59% | 22% |
| 9 | 9% | 0% | 13% | 33% | 45% |
| 10 | 2% | 0% | 9% | 61% | 28% |
| 11 | 0% | 4% | 22% | 37% | 37% |
| 12 | 0% | 4% | 30% | 39% | 27% |
| 13 | 0% | 2% | 39% | 46% | 13% |
| 14 | 2% | 20% | 46% | 30% | 2% |
| 15 | 0% | 2% | 28% | 54% | 16% |
| 16 | 0% | 4% | 28% | 52% | 16% |
| 17 | 4% | 28% | 46% | 20% | 2% |
| 18 | 0% | 0% | 41% | 48% | 11% |
| 19 | 2% | 2% | 41% | 37% | 18% |
| 20 | 2% | 0% | 7% | 61% | 30% |
| Total | 35% | 102% | 523% | 868% | 472% |
| Average | 1.75% | 5.1% | 26.15% | 43.4% | 23.600% |

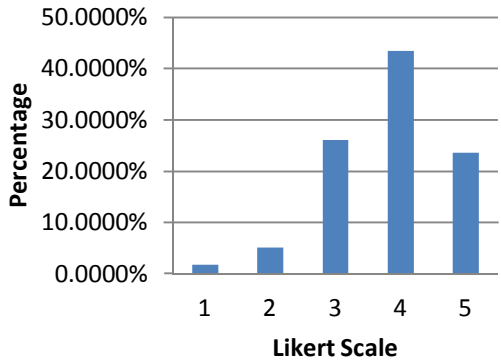

Figure 5. Graphical representation of 3 goals

**7. Conclusion**

　　　　The authors proposed a solution for one of the most problem related to components reuse. The proposed solution is improving the integrated component selection framework by adding the pliability metric that includes the software quality attributes in component selection. This proposed solution deals with selecting the appropriate components that satisfies the client's requirements. The questionnaire results reflect that the proposed solution improves the integrated component selection framework to select the suitable components.

**Corresponding Authors:**
Weam Gaoud Alghabban
Graduate Student of Information Technology
Faculty of Computing & Information Technology
King Abdulaziz University, Jeddah
E-mail: weam_ghabban@yahoo.com

Dr. Rizwan Jameel Qureshi
Assistant Professor of Information Technology,
Faculty of Computing & Information Technology
King Abdulaziz University, Jeddah
E-mail: anriz@hotmil.com

12/22/2013